\documentclass[aps,prd,onecolumn,groupedaddress,showpacs,nofootinbib,amssymb]{revtex4}
\usepackage[dvips]{graphicx}
\usepackage{amssymb}
\usepackage{amsmath}
\usepackage{graphicx,,color}
\usepackage{amsfonts}
\usepackage{bm}
\usepackage{cancel}
\usepackage{comment}

\newcommand\be{\begin{equation}}
\newcommand\ee{\end{equation}}

\newcommand{\ba}{\begin{eqnarray}}
\newcommand{\ea}{\end{eqnarray}}
\allowdisplaybreaks[4]

\allowdisplaybreaks[4]

\begin{document}

\title{Confronting Rainbow-deformed $f(R)$ Gravity with the ACT Data}
\author{S.D. Odintsov$^{1,2,3,4}$}\email{odintsov@ice.csic.es}
\author{V.K. Oikonomou,$^{4,5}$}\email{voikonomou@gapps.auth.gr;v.k.oikonomou1979@gmail.com}
\affiliation{$^{1)}$ Institute of Space Sciences (ICE, CSIC) C. Can Magrans s/n, 08193 Barcelona, Spain \\
$^{2)}$ ICREA, Passeig Luis Companys, 23, 08010 Barcelona, Spain\\
$^{3)}$Institut d'Estudis Espacials de Catalunya (IEEC), 08860
Castelldefels, Barcelona, Catalonia, Spain\\
$^{4)}$L.N. Gumilyov Eurasian National University - Astana,
010008, Kazakhstan \\
$^{5)}$Department of Physics, Aristotle University of
Thessaloniki, Thessaloniki 54124, Greece}

\tolerance=5000

\begin{abstract}
In this work we examine a theoretical scenario which combines two
fundamental theoretical proposals for the early Universe and a
possible evolution for the early post-inflationary epoch.
Specifically, we assume that the early Universe contains gravity's
rainbow effects on the spacetime and the inflationary Lagrangian
contains $R^2$ corrections or $f(R)$ gravity corrections in
general. In addition we assume that the era beyond the end of
inflation until the reheating temperature is reached, is a
kination era. Both theories, $R^2$ and gravity's rainbow emerge
from a quantum context so their effects should be first checked at
the theory which connects the quantum with the classical, hence
inflation. Spacetime is four dimensional and the effects of the
quantum theory could possibly be imprinted in the inflationary
Lagrangian and of course on the spacetime itself. For the
gravity's rainbow deformed Starobinsky model, both Lagrangian
quantum effects and spacetime quantum effects are combined. As we
show the resulting theory is compatible with the ACT data,
regarding the spectral index. We also consider power-law $f(R)$
gravity deformations and we show that in this case, the model is
viable without the need of extending the slow-roll era of
inflation.
\end{abstract}


\maketitle

\section{Introduction}

Inflation \cite{inflation1,inflation2,inflation3,inflation4}, is
the most well known theoretical proposal for the post-Planck
Universe. It solves many problems of the standard Big Bang theory,
and in addition it is the prominent theory known to connect the
quantum to the classical Universe. This remarkable theory is put
to test for nearly 3 decades now and still is in the focus of many
experiments focusing on astronomical data. It will be the main
focus of the Simons observatory \cite{SimonsObservatory:2019qwx},
and there is the possibility of detecting directly the B-mode of
the Cosmic Microwave Background (CMB) radiation. Apart from that,
there is also the indirect detection of the stochastic
gravitational wave background by the future gravitational wave
experiments
\cite{Hild:2010id,Baker:2019nia,Smith:2019wny,Crowder:2005nr,Smith:2016jqs,Seto:2001qf,Kawamura:2020pcg,Bull:2018lat,LISACosmologyWorkingGroup:2022jok}.
To date the existence of a stochastic gravitational wave
background has been confirmed by the NANOGrav
\cite{NANOGrav:2023gor} and other Pulsar Timing Array experiments
\cite{Antoniadis:2023ott,Reardon:2023gzh,Xu:2023wog}, however such
a background is rather unlikely to be generated solely by
inflation \cite{Vagnozzi:2023lwo,Oikonomou:2023qfz}. A remarkable
recent result for CMB physics was offered by the Atacama Cosmology
Telescope (ACT) data \cite{ACT:2025fju,ACT:2025tim}, which when
combined with the DESI data \cite{DESI:2024uvr} yield a scalar
spectral index which is in discordance with the Planck data
\cite{Planck:2018jri} for at least 2$\sigma$. Specifically the
spectral index of the scalar primordial perturbations is
constrained by the ACT data as follows,
\begin{equation}\label{act}
n_{\mathcal{S}}=0.9743 \pm
0.0034,\,\,\,\frac{\mathrm{d}n_{\mathcal{S}}}{\mathrm{d}\ln
k}=0.0062 \pm 0.0052\,\,\,(68\%) .
\end{equation}
Currently there are many proposals in the recent literature that
agree with the ACT data, for a mainstream of articles see for
example
\cite{Kallosh:2025rni,Gao:2025onc,Liu:2025qca,Yogesh:2025wak,Yi:2025dms,Peng:2025bws,Yin:2025rrs,Byrnes:2025kit,Wolf:2025ecy,Aoki:2025wld,Gao:2025viy,Zahoor:2025nuq,Ferreira:2025lrd,Mohammadi:2025gbu,Choudhury:2025vso,Odintsov:2025wai,Q:2025ycf,Zhu:2025twm,Kouniatalis:2025orn,Hai:2025wvs,Dioguardi:2025vci,Yuennan:2025kde,Kuralkar:2025zxr,Kuralkar:2025hoz,Aoki:2025ywt}.
Although we share the approach of \cite{Ferreira:2025lrd}, so it
is quite early to be certain about the ACT data, which must be
used with skepticism. Note that apart from the ACT constraints on
the scalar spectral index, in this work we shall consider the
Planck updates on the tensor-to-scalar ratio \cite{BICEP:2021xfz},
\begin{equation}\label{planck}
r<0.036
\end{equation}
at $95\%$ confidence.

In this article we aim to provide another theory, which is
theoretically well motivated and can be compatible with the ACT
data. Specifically we shall consider the kinetic extended
gravity's rainbow deformed Starobinsky model. The analysis of such
model was thoroughly performed in Refs.
\cite{Waeming:2020rir,Chatrabhuti:2015mws,Channuie:2019kus} and
this subject is quite popular in the literature
\cite{Feng:2016zsj,Hendi:2016hbe,Deng:2017umx,Khodadi:2016aop,Khodadi:2016bcx,Rudra:2016alu,Ashour:2016cay,Garattini:2012ec,Garattini:2014rwa,Hendi:2016tiy,Hendi:2017vgo,Heydarzade:2017rpb,Hendi:2016oxk,Chatrabhuti:2015mws,Codello:2014sua}.
This theoretical context is very well motivated for two reasons:
firstly, $R^2$ terms are among the quantum correction terms of the
standard scalar field action and also the gravity's rainbow idea,
is a remarkable idea which states that the spacetime metric might
be an energy dependent quantity, similar to the rainbow effect in
optics, where the diffractive index depends on the wavenumber of
the light beam. For the first motivation we mentioned, the most
general scalar field Lagrangian with two derivatives in dimension
four is,
\begin{equation}\label{generalscalarfieldaction}
\mathcal{S}_{\varphi}=\int
\mathrm{d}^4x\sqrt{-g}\left(\frac{1}{2}Z(\varphi)g^{\mu
\nu}\partial_{\mu}\varphi
\partial_{\nu}\varphi+\mathcal{V}(\varphi)+h(\varphi)\mathcal{R}
\right)\, .
\end{equation}
The one loop quantum corrections consistent with diffeomorphism
invariance, to the above action are \cite{Codello:2015mba},
\begin{align}\label{quantumaction}
&\mathcal{S}_{eff}=\int
\mathrm{d}^4x\sqrt{-g}\Big{(}\Lambda_1+\Lambda_2
\mathcal{R}+\Lambda_3\mathcal{R}^2+\Lambda_4 \mathcal{R}_{\mu
\nu}\mathcal{R}^{\mu \nu}+\Lambda_5 \mathcal{R}_{\mu \nu \alpha
\beta}\mathcal{R}^{\mu \nu \alpha \beta}+\Lambda_6 \square
\mathcal{R}\\ \notag &
+\Lambda_7\mathcal{R}\square\mathcal{R}+\Lambda_8 \mathcal{R}_{\mu
\nu}\square \mathcal{R}^{\mu
\nu}+\Lambda_9\mathcal{R}^3+\mathcal{O}(\partial^8)+...\Big{)}\, ,
\end{align}
with the parameters $\Lambda_i$, $i=1,2,...,6$ being dimensionful
constants. Thus the $R^2$ term emerges directly from a quantum
context and is thus a quantum effect on the inflationary
Lagrangian. Note the presence of higher derivatives is present,
see for example Ref. \cite{Myrzakulov:2014hca} for inflationary
analysis of this sort. On the other hand, the gravity's rainbow
theory is a quantum gravity deformation of spacetime itself. Such
quantum gravity effects modify the photon or other particles
dispersion relations and can be verified by the Cherenkov Array
Telescope \cite{Cherenkov}. In the context of gravity's rainbow,
the standard Lorentz invariant dispersion relation
$\epsilon^{2}-p^{2}=m^{2}$ is replaced by $\epsilon^{2}{\tilde
f}^{2}(\epsilon) - p^{2}{\tilde g}^{2}(\epsilon)=m^{2}$ with
${\tilde f}(\epsilon)$ and ${\tilde g}(\epsilon)$ being the
rainbow functions. The spacetime itself is modified in the context
of gravity's rainbow and in a cosmological context, the
Friedmann-Robertson-Walker metric is modified as follows,
\begin{eqnarray}
\mathrm{d}s^2(\epsilon)
=-\frac{dt^2}{\tilde{f}^2(\epsilon)}+\frac{a^2(t)}{\tilde{g}^2(\epsilon)}
\delta_{ij}{dx}^i{dx}^j\,.\label{FRW}
\end{eqnarray}
We shall consider the case $\tilde{g}=1$ for simplicity and adopt
the functional form for the function $\tilde{f}$ used in Ref.
\cite{Waeming:2020rir,Chatrabhuti:2015mws,Channuie:2019kus}, and
we shall also adopt their notation for simplicity and easy
referral.

The gravity's rainbow deformed Starobinsky model is a well
motivated theoretical context since it combines both a quantum
corrected Lagrangian and low-energy quantum effects of the quantum
gravity era on the spacetime itself. The best test for this
context is the inflationary era, which is a theory connecting the
quantum era of our Universe and the classical Universe, where the
spacetime is four dimensional and the effective Lagrangian of the
inflationary Universe may contain remnants of the quantum era.
Under the assumption that the post-inflationary era until the
reheating temperature is reached, is governed by a kination
regime, we show that gravity's rainbow deformed Starobinsky model
is compatible with the ACT data. Kination deformations of the
inflationary era \cite{Oikonomou:2022tux} are motivated by kinetic
axion models \cite{Co:2019jts,Co:2020dya,Barman:2021rdr}.

\section{General Setup: $f(R)$ Gravity's Rainbow and Cosmological Perturbations}

In this section we shall briefly review how the $f(R)$ gravity
framework is affected by the spacetime deformation of Eq.
(\ref{FRW}). We shall adopt the notation of Ref.
\cite{Waeming:2020rir,Chatrabhuti:2015mws,Channuie:2019kus},
slightly changing only the notation for $\epsilon_2$ and
$\epsilon_3$ to $\epsilon_3$ and $\epsilon_4$ in order to comply
with the standard literature of $f(R)$ gravity
\cite{reviews1,reviews2}. The $f(R)$ gravity gravitational field
equations in vacuum are, \ba F(R)R_{\mu\nu}(g) -
\frac{1}{2}f(R)g_{\mu\nu}-\nabla_{\mu}\nabla_{\nu}F(R)+g_{\mu\nu}\Box
F(R) = 0\,, \label{eom} \ea with $F(R)=\partial f(R)/\partial R$.
For the metric (\ref{FRW}) we get the following field equations,
\begin{eqnarray}
3 \left(FH^2+H\dot{F} \right) -6FH\frac{  \dot{\tilde{g}}}{\tilde{g}} +3 F\frac{ \dot{\tilde{g}}^2}{\tilde{g}^2}+\dot{F}\frac{ \dot{\tilde{f}}}{\tilde{f}}-3 \dot{F}\frac{ \dot{\tilde{g}}}{\tilde{g}}=\frac{F R-f(R)}{2 \tilde{f}^2}+\frac{\kappa ^2 \rho }{\tilde{f}^2}\,,\label{Hdd}
\end{eqnarray}
\begin{eqnarray}
&&3 F H^2-3 \dot{F} H+3 F \dot{H}+3 F H\frac{ \dot{\tilde{f}}}{\tilde{f}}-\dot{F}\frac{ \dot{\tilde{f}}}{\tilde{f}}-4 F\frac{ \dot{\tilde{g}}^2}{\tilde{g}^4}+6 F H\frac{ \dot{\tilde{g}}}{\tilde{g}^3}-3 \dot{F}\frac{ \dot{\tilde{g}}}{\tilde{g}^3}+F\frac{ \ddot{\tilde{g}}}{\tilde{g}^3}+F\frac{\dot{\tilde{f}} }{\tilde{f}}\frac{\dot{\tilde{g}}}{\tilde{g}^3}-3 F H^2\frac{1}{\tilde{g}^2}\nonumber\\&&+2 \dot{F} H\frac{1}{\tilde{g}^2}+\frac{\ddot{F}}{\tilde{g}^2}-F \dot{H}\frac{1}{\tilde{g}^2}+6 F\frac{ \dot{\tilde{g}}^2}{\tilde{g}^2}-F H\frac{ \dot{\tilde{f}}}{\tilde{f} }\frac{1}{\tilde{g}^2}+\dot{F}\frac{ \dot{\tilde{f}}}{\tilde{f}}\frac{1}{\tilde{g}^2}-6 F H\frac{ \dot{\tilde{g}}}{\tilde{g}}+3 \dot{F}\frac{ \dot{\tilde{g}}}{\tilde{g}}-3 F\frac{ \ddot{\tilde{g}}}{\tilde{g}}-3 F\frac{\dot{\tilde{f}} }{\tilde{f} }\frac{\dot{\tilde{g}}}{\tilde{g}} \nonumber\\&&-\frac{f(R) \left(\tilde{g}-1\right) \left(\tilde{g}+1\right)}{2 \tilde{f}^2 \tilde{g}^2}=-\frac{\kappa ^2 \left(\rho  \tilde{g}^2+P\right)}{\tilde{f}^2 \tilde{g}^2}\,,\label{ijcom}
\end{eqnarray}
and notice the presence of the terms containing the rainbow
functions $\tilde{g}$ and $\tilde{f}$. From now on we shall
consider only the case $\tilde{g}=1$, $\tilde{f}\neq 1$. Following
well known steps in the literature \cite{Hwang:2005hb}, the
authors of \cite{Waeming:2020rir} derived the cosmological
perturbations for the gravity's rainbow deformed $f(R)$ gravity,
which take the following form,
\begin{eqnarray}\label{a1}
    ds^2 &=&-\frac{(1+2\alpha )}{\tilde{f}^2(\epsilon)}dt^2-\frac{2 a(t) \left(\partial_i\beta-S_i\right)}{\tilde{f}(\epsilon )}dt dx^i\nonumber\\&&+a^2(t)  \left(\delta _{ij}+2 \psi  \delta _{ij}+2 \partial_i\partial_j \gamma +2\partial_j F_i+h_{ij}\right)dx^i dx^j\,,
\end{eqnarray}
with $\alpha,\beta,\psi,\gamma$ being the scalar perturbations,
$S_i,F_i$ being vector perturbations and $h_ij$ are the tensor
perturbations. The slow-roll indices for inflation are defined as
\cite{Hwang:2005hb,reviews1},
\begin{eqnarray}
 \epsilon_1 = -\frac{\dot{H}}{H^2},  \ \ \epsilon_3 = \frac{\dot{F}}{2HF}, \ \ \epsilon_4 = \frac{\dot{E}}{2HE}\ ,
\end{eqnarray}\label{a20}
with $E \equiv 3\dot{F}^2/2\kappa^2$. A simple form for the
rainbow function $\tilde{f}$ is chosen and specifically the
following
\cite{Waeming:2020rir,Chatrabhuti:2015mws,Channuie:2019kus},
\begin{equation}\label{rainbowfunction}
\tilde{f}^2=1+ \left(\frac{H}{M}\right)^{2\lambda}\, ,
\end{equation}
where $M$ is the parameter also appearing in the Starobinsky
model,
\begin{equation}\label{starobinskymodel}
f(R)=R+\frac{R^2}{6M^2}\, .
\end{equation}
Note that during inflation we have, $\tilde{f}\sim
\left(\frac{H}{M}\right)^{2\lambda}$ while at the post
inflationary era, where the curvature of the Universe drops, we
have $\tilde{f}\simeq 1$. The spectral index of the scalar
perturbations is written in terms of the slow-roll parameters as
\cite{Waeming:2020rir,Chatrabhuti:2015mws,Channuie:2019kus},
\begin{eqnarray}
n_{\mathcal{S}} - 1 =
\sqrt{\frac{1}{4}+\frac{(1+\epsilon_1-\epsilon_3+\epsilon_4)(2-\lambda)\epsilon_1-\epsilon_3+\epsilon_4)}{(1-(1+\lambda)\epsilon_1)^2}}
\, , \label{a31}
\end{eqnarray}
without however taking the slow-roll expansion approximation, in
difference with \cite{Hwang:2005hb}, while the tensor spectral
index is given by the following relation \cite{Hwang:2005hb},
\begin{equation}\label{tensorspectralindexini}
n_T=3-2\nu_t
\end{equation}
where,
\begin{eqnarray}
 \nu^2_t = \frac{1}{4} + \frac{(1+\epsilon_3)(2-(1+\lambda)\epsilon_1+\epsilon_3)}{(1-(1+\lambda)\epsilon_1)^2} \, . \label{a38}
\end{eqnarray}
 Finally, the tensor-to-scalar ratio is \cite{Odintsov:2020thl}\footnote{Note there is a typo in \cite{Waeming:2020rir} also appearing in other texts in the literature},
\begin{eqnarray}
 r\simeq \frac{8\kappa^2Q_s}{F} \, , \label{a43}
\end{eqnarray}
where,
\begin{eqnarray}
 Q_s = \frac{3\dot{F}^2}{2\kappa^2F(H+\dot{F}/2F)^2} \ . \label{a18}
\end{eqnarray}
so without taking the slow-roll expansion we get,
\begin{eqnarray}
 r = \frac{48 \epsilon_3^2}{(\epsilon_3+1)^2}\ , \label{a44}
\end{eqnarray}
which is slightly different from Ref. \cite{Waeming:2020rir}, see
\cite{reviews2} for details and also \cite{Odintsov:2020thl}.
However, for the sake of completeness we shall also consider the
case considered in \cite{Waeming:2020rir} with,
\begin{equation}\label{a44a}
r=48\epsilon_3^2\, .
\end{equation}
Now the aim is to find the analytic form of the slow-roll indices
$\epsilon_1$, $\epsilon_3$ and $\epsilon_4$ for the Starobinsky
model (\ref{starobinskymodel}). Assuming a slow-roll inflationary
era, the slow-roll index $\epsilon_1$ is derived from the field
equations \cite{Waeming:2020rir},
\begin{equation}\label{epsilon1f}
\epsilon_1\simeq \frac{1}{2 N}\, ,
\end{equation}
which as we mentioned is easily obtained from the $R^2$ field
equations. Also from the Raychaudhuri equation for the gravity's
rainbow deformed Starobinsky model, we have,
\begin{eqnarray}
    \ddot{F}-H\dot{F} +2 F \dot{H}+2 F H\frac{\dot{\tilde{f}}}{\tilde{f}}=0.\label{4.28}
\end{eqnarray}
Using $\tilde{f}\approx (H/M)^\lambda$, we have
\begin{eqnarray}
    \frac{\dot{\tilde{f}}}{\tilde{f}}=\frac{\lambda \dot{H}}{H}.
\end{eqnarray}
Thus, Eq.(\ref{4.28}) can be rewritten as
\begin{eqnarray}
    \ddot{F}-H\dot{F} +2(1+\lambda) F \dot{H}=0.\label{4.30}
\end{eqnarray}
Dividing Eq.(\ref{4.30}) by $2H^2F$, then it reduces to
\begin{eqnarray}
    \frac{\ddot{F}}{2H^2F}-\epsilon_2 -(1+\lambda)\epsilon_1=0\, ,\label{4.31x}
\end{eqnarray}
so we get,
\begin{eqnarray}
    \epsilon_3(\epsilon_4-1)-(1+\lambda)\epsilon_1=0\, .\label{4.31}
\end{eqnarray}
hence,
\begin{equation}\label{epsilon41}
\epsilon_1=\frac{\epsilon_3 (1+\lambda)}{\epsilon_4-1}\, .
\end{equation}
Now assuming a slow-roll era, we have the approximation,
\begin{equation}\label{curva1}
R\simeq \frac{12 H(t)^{2\lambda +2}}{M^{2\lambda }}\, ,
\end{equation}
and from it, by omitting terms $\sim \ddot{H}$, we get,
\begin{equation}\label{rdot}
\dot{R}\simeq 24 (\lambda +1) M^{-2\lambda } H^{2\lambda +1}
\dot{H}
\end{equation}
and
\begin{equation}\label{rddot}
\ddot{R}\simeq 24 (\lambda +1) (2 \lambda +1) M^{-2 \lambda }
H(t)^{2 \lambda } \dot{H}^2\, ,
\end{equation}
and since for the Starobinsky model we have,
\begin{equation}\label{starobiskyespilon4}
\epsilon_4=\frac{\ddot{R}}{H\dot{R}}\, ,
\end{equation}
by combining Eqs. (\ref{rdot}), (\ref{rddot}) and
(\ref{starobiskyespilon4}) we get,
\begin{equation}\label{epsilon4f}
\epsilon_4=-(1 + 2\lambda)\epsilon_1\, .
\end{equation}
Now equipped Eqs. (\ref{a31}), (\ref{a44}), (\ref{epsilon1f}),
(\ref{epsilon41}) and  (\ref{epsilon4f}), we can proceed and
confront the model with the ACT data (\ref{act}), using also Eq.
(\ref{a44a}) for the tensor-to-scalar ratio for completeness. A
thorough analysis reveals that there is no positive value for
$\lambda$ that can yield an inflationary phenomenology compatible
with the ACT data for $N\sim 60$. One needs larger $e$-fold
numbers to achieve compatibility. Specifically, for the
Starobinsky model ($\lambda=0$) one needs at least $N\sim 69$. But
this kind of extension of the inflationary era is rather
unmotivated, even after the inflationary era the equation of state
(EoS) parameter is not $w=1/3$ but generally $w$. Let us see why,
and let us consider the $e$-foldings number for a general
primordial scalar mode with a wavenumber $k$, which just became
superhorizon at the beginning of inflation, which is equal to
\cite{Adshead:2010mc,Oikonomou:2022tux},
\begin{equation}\label{generalefoldingsnumber}
\frac{a_kH_k}{a_0H_0}=e^{-N}\frac{H_ka_{end}}{a_{reh}H_{reh}}\frac{H_{reh}a_{reh}}{a_{eq}H_{eq}}\frac{H_{eq}a_{eq}}{a_{0}H_{0}}\,
,
\end{equation}
where $a_k$ and $H_k$ are the scale factor and the Hubble rate at
exactly the time instance at which the mode $k$ became
superhorizon exactly at the beginning of inflation (that is at
first horizon crossing).
\begin{figure}
\centering
\includegraphics[width=25pc]{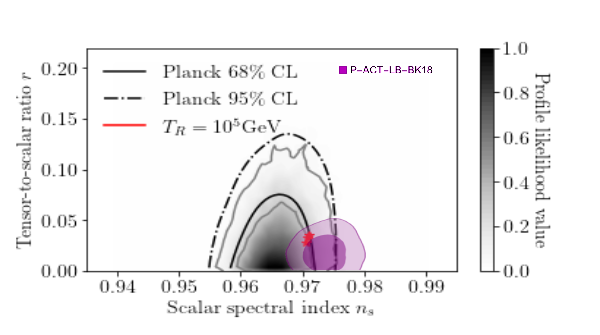}
\includegraphics[width=25pc]{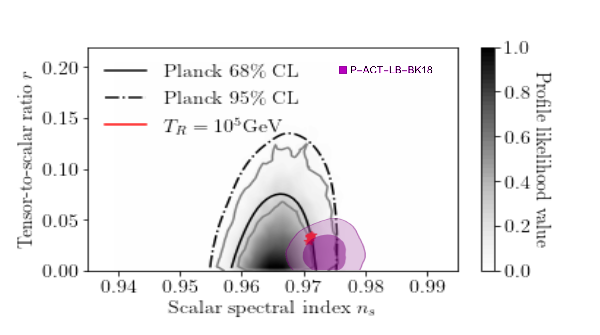}
\caption{Marginalized curves of the Planck 2018 data and the
kination extended gravity's rainbow deformed Starobinsky model
confronted with the ACT, the Planck 2018 data and the updated
Planck constraint on the tensor-to-scalar ratio for an
intermediate reheating temperature $T_R=10^5\,$GeV and the
tensor-to-scalar ratio given by Eq. (\ref{a44}) (top plot) and by
Eq. (\ref{a44a}) (bottom plot). The pivot scale is the CMB pivot
scale $k_*=0.05$Mpc$^{-1}$ and also we used $N=65.2118$ and
$\lambda=[2.2, 2.644]$ in the plots.}\label{plot1}
\end{figure}
Also $a_{end}$ denotes the scale factor at the end of inflation,
and also $a_{reh}$ and $H_{reh}$ stand for the scale factor and
the Hubble rate at the end of the reheating era. In addition,
$a_{eq}$ and $H_{eq}$ stand for the scale factor and the Hubble
rate at the matter-radiation equality, and in addition $a_0$ and
$H_0$ are the present day scale factor and the Hubble rate. If at
the end of inflation, the total EoS parameter is $w$, instead of
the radiation domination value $w=1/3$, we get,
\begin{equation}\label{aux1}
\ln \left(\frac{a_{end}H_{end}}{a_{reh}H_{reh}}
\right)=-\frac{1+3w}{6(1+w)}\ln
\left(\frac{\rho_{reh}}{\rho_{end}} \right)\, ,
\end{equation}
where $H_{end}$ denotes the Hubble rate at the end of inflation,
and $\rho_{end}$ and $\rho_{reh}$ denote the total energy
densities of the Universe at the end of inflation and at the end
of the reheating era, where we assumed that the EoS parameter at
the end of inflation and at the end of the reheating era is
constant. Further working out Eq. (\ref{aux1}) we get
\cite{Adshead:2010mc},
\begin{equation}\label{efoldingsmainrelation}
N=56.12-\ln \left( \frac{k}{k_*}\right)+\frac{1}{3(1+w)}\ln \left(
\frac{2}{3}\right)+\ln \left(
\frac{\rho_k^{1/4}}{\rho_{end}^{1/4}}\right)+\frac{1-3w}{3(1+w)}\ln
\left( \frac{\rho_{reh}^{1/4}}{\rho_{end}^{1/4}}\right)+\ln \left(
\frac{\rho_k^{1/4}}{10^{16}\mathrm{GeV}}\right)\, ,
\end{equation}
where $\rho_k$ is the Universe's total energy density when
inflation starts, when the mode $k$ became superhorizon. We take
the pivot scale $k_*$ to be $k_*=0.05$Mpc$^{-1}$ and we assume
that the degrees of freedom of particles $g_*$ is nearly constant.
Now in a model independent way, assuming a high scale inflation
$\rho_k^{1/4}\sim 10^{16}\,$GeV, and assuming a kination era after
the inflationary era, we get the $e$-foldings numbers presented in
Table \ref{table1} for three reheating temperatures, a high
$T_R=10^{12}\,$GeV, an intermediate reheating temperature
$T_R=10^{5}\,$GeV, and a low reheating temperature
$T_R=10^{2}\,$GeV.
\begin{table}[h!]
  \begin{center}
    \label{table1}
    \begin{tabular}{|r|r|r|r|}
     \hline
      \textbf{$e$-foldings number} & \textbf{$T_R=10^{12}\,$GeV} & \textbf{$T_R=10^{5}\,$GeV} & \textbf{$T_R=10^{2}\,$GeV} \\
           \hline
           $e$-foldings number $N$ & 59.8391 & 65.2118 & 67.5143\\ \hline
      \end{tabular}
  \end{center}
      \caption{\emph{\textbf{The $e$-foldings number for a general kinetically extended inflationary era for various reheating temperatures. }}}
\end{table}
Of course, there is the possibility of lower reheating
temperatures, for example \cite{Hasegawa:2019jsa}, but we did not
consider this possibility. In our analysis we considered a high
scale inflationary era with $V_k^{1/4}\sim 1.73 \times
10^{16}\,$GeV and $V_{end}^{1/4}\sim 0.75 V_k^{1/4}$ which is
observationally correct and motivated. In addition we assumed that
$g_*$ is constant across the temperature range and in addition we
neglected entropy production during reheating era.

From the above considerations, it is obvious that even a kinetic
extended inflationary era cannot make the Starobinsky model
compatible with the ACT data. However, the gravity's rainbow
deformed Starobinsky model can be compatible with the Planck data,
if the inflationary era is extended due to the occurrence of a
kination era. For example an axion field can be responsible for
that extension \cite{Oikonomou:2022tux}. Let us consider the
values of $\lambda$ and $e$-foldings number that can achieve such
a scenario, for gravity's rainbow deformed Starobinsky model, for
the three reheating temperature cases presented in Table
\ref{table1}. Let us consider first the low reheating temperature
case, in which case $N=67.5143$, so for the tensor-to-scalar ratio
being given by Eq. (\ref{a44}) one gets that $\lambda$ must be at
most $\lambda_{max}=2.773$ in order to have compatibility with the
ACT data (\ref{act}), with the corresponding $n_{\mathcal{S}}$
being $n_{\mathcal{S}}=0.972183$ and $r=0.0359837$. For the
tensor-to-scalar ratio being given by Eq. (\ref{a44a}) one gets
that $\lambda$ must be in this case, at most $\lambda_{max}=2.883$
in order to have compatibility with the ACT data (\ref{act}), with
the corresponding $n_{\mathcal{S}}$ being
$n_{\mathcal{S}}=0.972325$ and $r=0.0359962$.

Now let us consider the intermediate reheating temperature case,
in which case $N=65.2118$, so for the tensor-to-scalar ratio being
given by Eq. (\ref{a44}) one gets in this case that $\lambda$ must
be at most $\lambda_{max}=2.644$ in order to have compatibility
with the ACT data (\ref{act}), with the corresponding
$n_{\mathcal{S}}$ being $n_{\mathcal{S}}=0.971093$ and
$r=0.0359959$. For the tensor-to-scalar ratio being given by Eq.
(\ref{a44a}) one gets that $\lambda$ must be in this case, at most
$\lambda_{max}=2.745$ in order to have compatibility with the ACT
data (\ref{act}), with the corresponding $n_{\mathcal{S}}$ being
$n_{\mathcal{S}}=0.971227$ and $r=0.035913$. Regarding the high
reheating temperature case, there is no positive value of
$\lambda$ that can render the model compatible with the ACT data.

\begin{figure}
\centering
\includegraphics[width=25pc]{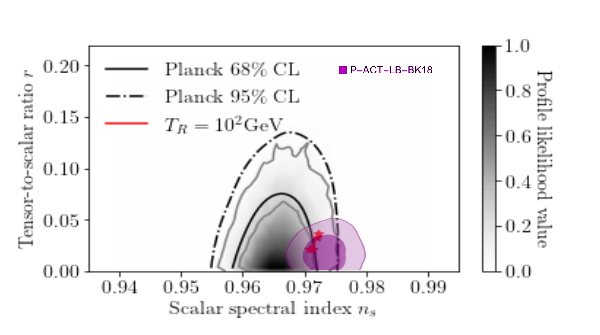}
\includegraphics[width=25pc]{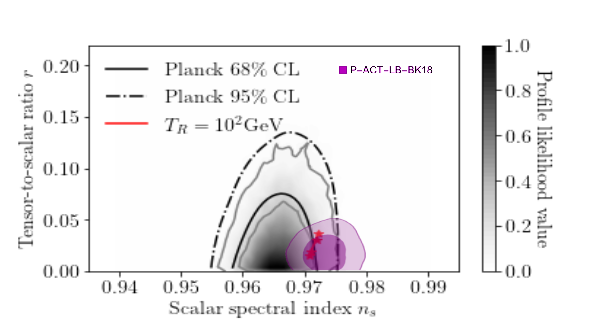}
\caption{Marginalized curves of the Planck 2018 data and the
kination extended gravity's rainbow deformed Starobinsky model
confronted with the ACT, the Planck 2018 data and the updated
Planck constraint on the tensor-to-scalar ratio for a low
reheating temperature $T_R=10^2\,$GeV and the tensor-to-scalar
ratio given by Eq. (\ref{a44}) (top plot) and by Eq. (\ref{a44a})
(bottom plot). The pivot scale is the CMB pivot scale
$k_*=0.05$Mpc$^{-1}$ and also we used $N=67.5143$ and
$\lambda=[1.473,2.773]$ in the plots.}\label{plot2}
\end{figure}

\begin{figure}
\centering
\includegraphics[width=25pc]{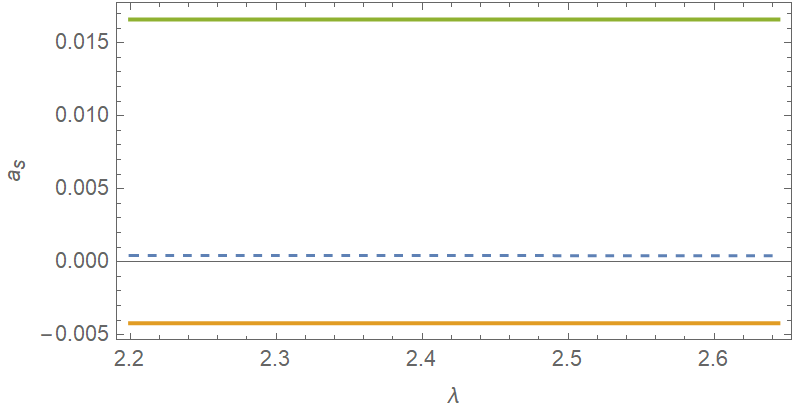}
\caption{The running of the spectral index $a_s$ for the
intermediate reheating case (dashed curve), versus the $95\%$
confidence ACT data on the running of the spectral index (thick
curves), for $\lambda=[2.2,2.644]$.  }\label{runplot1}
\end{figure}

In order to have a clearer idea of the viability of the kination
extended gravity's rainbow deformed Starobinsky model, in Figs.
\ref{plot1} and \ref{plot2} we present the confrontation of the
kination extended gravity's rainbow deformed Starobinsky model
with the ACT and Planck 2018 data \cite{Planck:2018jri}, for
intermediate and low reheating temperatures respectively, using
Eq. (\ref{a44}) for the tensor-to-scalar ratio.
As it can be seen, both the low and intermediate reheating
temperature cases are marginally, but still, compatible with the
ACT data and specifically the lower limits of the data regarding
the scalar spectral index. This is an interesting scenario which
we deemed reportable. Furthermore, let us consider the running of
the spectral index in the present framework. The running of the
spectral index is defined in the following way,
\begin{equation}\label{runningdef}
a_s=\frac{\mathrm{d} n_{\mathcal{S}}}{\mathrm{d} \ln k}\, ,
\end{equation}
with $k$ being the comoving wavenumber of a primordial mode. We
can recast $a_s$ in the following way,
\begin{equation}\label{runningdef1}
a_s=\frac{\mathrm{d} n_{\mathcal{S}}}{\mathrm{d} \ln
k}=\frac{\mathrm{d}
n_{\mathcal{S}}}{\mathrm{d}N}\frac{\mathrm{d}N}{\mathrm{d} \ln
k}\, ,
\end{equation}
with $N$ being the $e$-foldings number. Furthermore, by using the
formula $\frac{\mathrm{d}N}{\mathrm{d} \ln
k}=\frac{1}{1-\epsilon_1}$, we obtain the final expression for the
running of the spectral index $a_s$, which is,
\begin{equation}\label{runningdefmainfinal}
a_s=\frac{1}{1-\epsilon_1}\frac{\mathrm{d}
n_{\mathcal{S}}}{\mathrm{d}N}\, .
\end{equation}
For the model at hand, we have,
\begin{equation}\label{runninganalyticalstarobinsky}
a_{s}=\frac{8 N \sqrt{\frac{\left(10 \lambda ^2+7 \lambda -6
\lambda  N-4 N (3 N+2)+1\right)^2}{(\lambda -2 N+1)^2 (2 \lambda
+2 N+1)^2}} \left(\lambda  (\lambda  (5-2 \lambda )+7)+8 N^2+4
\left(-2 \lambda ^2+\lambda +1\right) N+2\right)}{(2 N-1)
(-\lambda +2 N-1) (2 \lambda +2 N+1) \left(-10 \lambda ^2-7
\lambda +6 \lambda  N+4 N (3 N+2)-1\right)}\, .
\end{equation}
Now we can confront the running of the spectral index for the
current model with the ACT data. In Fig. \ref{runplot1} we present
the results of our analysis and the ACT constraints on the running
of the spectral index, where we took the $95\%$ confidence level
results which constraint the running of the spectral index to be
in the range: $a_s=[-0.0042,0.0166]$. As it can be seen, the model
is well fitted within the $95\%$ confidence level constraints of
ACT regarding the running of the spectral index. The result is at
$1.2\sigma$ conflict with the $68\%$ constraints, so
insignificant.
Furthermore we need to determine the scale of the parameter $M$
which solely affect the amplitude of the scalar perturbations and
the result is $M\sim (0.34-1.70)\times 10^{14}\,$GeV.

\subsection{Power-law $f(R)$ Gravity Case}

\begin{figure}
\centering
\includegraphics[width=28pc]{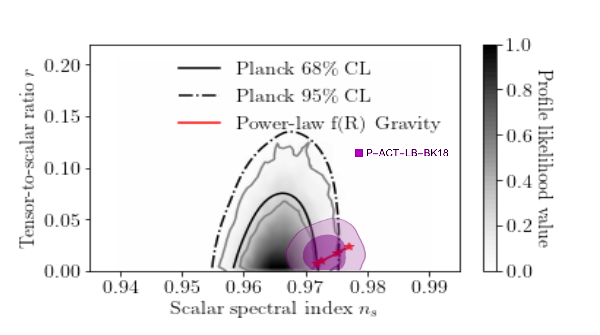}
\caption{Marginalized curves of the Planck 2018 data and the
power-law $f(R)$ gravity model confronted with the ACT, the Planck
2018 data, for $n=2.1$, $N\sim 60$ and $\lambda$ chosen in the
range $\lambda=[0.8-1.7]$. The pivot scale is the CMB pivot scale
$k_*=0.05$Mpc$^{-1}$.}\label{plot3}
\end{figure}

\begin{figure}
\centering
\includegraphics[width=25pc]{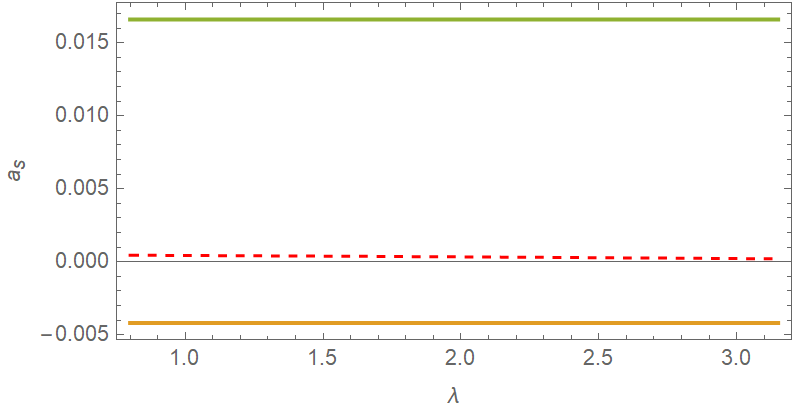}
\caption{The running of the spectral index $a_s$ for the power-law
$f(R)$ model (dashed curve), versus the $95\%$ confidence ACT data
on the running of the spectral index (thick curves), for
$\lambda=[0.8, 3.15]$ and $n=2.1$. }\label{runplot2}
\end{figure}

Now let us consider another interesting scenario, also considered
in Ref. \cite{Waeming:2020rir}, namely the power-law $f(R)$
gravity case. In this scenario, the $f(R)$ gravity has the form,
\begin{equation}\label{powerlawfr}
f(R)=R+\alpha M^2\left(\frac{R}{M^2} \right)^n\, ,
\end{equation}
so it is a power-law $f(R)$ gravity
\cite{Capozziello:2002rd,Sebastiani:2013eqa}. In principle,
following same technique we develop here, one can make
ACT-compatible any $f(R)$ model considered in Ref.
\cite{Sebastiani:2013eqa}. Now in this case, the field equations
yield again \cite{Waeming:2020rir},
\begin{equation}\label{epsilon1fpowerl}
\epsilon_1\simeq \frac{1}{2 N}\, ,
\end{equation}
and a simple analysis as in the $R^2$ yields,
\begin{eqnarray}\label{a31powlaw}
n_{\mathcal{S}} - 1 =
\sqrt{\frac{1}{4}+\frac{(1+\epsilon_1-\epsilon_3+\epsilon_4)(2-\lambda)\epsilon_1-\epsilon_3+\epsilon_4)}{(1-(1+\lambda)\epsilon_1)^2}}
\, ,
\end{eqnarray}
and also,
\begin{eqnarray}\label{a44powlaw}
 r = \frac{48 \epsilon_3^2}{(\epsilon_3+1)^2}\, ,
\end{eqnarray}
and in addition, without any change from the previously studied
Starobinsky case,
\begin{equation}\label{epsilon41powlaw}
\epsilon_3=\frac{\epsilon_3 (1+\lambda)}{\epsilon_4-1}\, .
\end{equation}
The only change between the power-law $f(R)$ gravity model and the
Starobinsky model at the level of the slow-roll indices is sourced
at the slow-roll index $\epsilon_4$. Specifically, for the
power-law $f(R)$ gravity model (\ref{powerlawfr}) we get from the
definition,
\begin{equation}\label{epsilon4powerlaw}
\epsilon_4=\frac{F''\dot{R}}{H\dot{R}}+\frac{\ddot{R}}{H\dot{R}}\,
,
\end{equation}
where recall $F=\frac{\partial f}{\partial R}$, and after some
simple calculations, we get,
\begin{equation}\label{finalepsilon4}
\epsilon_4=-(n-2)(2+2\lambda)\epsilon_1-(1+2\lambda)\epsilon_1\, .
\end{equation}
Now this case is quite interesting because it is not necessary to
extend the inflationary era and yields viable and ACT compatible
results for $N\sim 60$. For example if $\lambda=1.7$ and $n=2.1$
we get, $n_{\mathcal{S}}=0.97692$ and $r=0.023417$. Now to have a
clear picture of the viability of the model and its compatibility
with the ACT data, in Fig. \ref{plot3} we present the
confrontation of the power-law $f(R)$ gravity model
(\ref{powerlawfr}) with the ACT and Planck 2018 data
\cite{Planck:2018jri}, for $n=2.1$, $N\sim 60$ and $\lambda$
chosen in the range $\lambda=[0.8-1.7]$.
As it can be seen in Fig. \ref{plot3}, the power-law $f(R)$
gravity model is well fitted within the ACT and updated Planck
data. In addition, the model predicts a low running of the
spectral index, consistent with the $95\%$ constraints of the ACT
data on the running of the spectral index. Indeed, in Fig.
\ref{runplot2} we present the results of our analysis and the ACT
constraints on the running of the spectral index. As it can be
seen, this model too is viable in this aspect, so we need to
report this.

\section{Conclusions}

In this work we examined a theoretical proposal for a deformation
of the Starobinsky model which provides results that are
compatible with the ACT data. Specifically, our scenario combined
two fundamental theoretical proposals for the early epoch of our
Universe and a possible evolution for the early post-inflationary
epoch. The inflationary epoch is a classical regime of our
Universe, and we assumed that the quantum epoch, the so-called
post-Planck epoch had its imprint on the inflationary epoch in a
two fold way, firstly it affected the inflationary Lagrangian
adding an $R^2$ term in it, and secondly the spacetime itself
contains ultraviolet (UV) corrections in the form of gravity's
rainbow deformations of the metric. Furthermore we assumed that
the early post-inflationary epoch is kination dominated, until the
Universe reaches its reheating temperature. The latter effect
affects the duration of the inflationary epoch and as we showed,
for intermediate and low reheating temperatures, the $e$-foldings
number of the inflationary epoch is extended. As we showed, the
resulting theory is marginally compatible with the ACT data. Our
scenario is based on two quite possible UV completions of our
classical Universe, which can be observed in the future
independently. Thus the framework is quite well motivated, and the
Starobinsky model may thus be fitted with the ACT data, not well
fitted, but still fitted. However as we demonstrated, the rainbow
extended power-law $f(R)$ gravity model can be compatible with the
ACT data without extending the inflationary era. In addition,
further considerations of our framework can be performed in the
lines of geometric realizations or gravitational realizations of
rainbow gravity
\cite{Kouniatalis:2024gnr,Anagnostopoulos:2022gej,Dimakis:2022rkd,Anagnostopoulos:2021ydo,Garattini:2014rwa},
and also one should take into account plateau potentials with
ultra-suppressed tensor-to-scalar ratio
\cite{Kouniatalis:2025orn}.

\section*{Acknowledgments}

This work was partially supported by the program Unidad de
Excelencia Maria de Maeztu CEX2020-001058-M, Spain (S.D.
Odintsov). This research has been funded by the Committee of
Science of the Ministry of Education and Science of the Republic
of Kazakhstan (Grant No. AP26194585) (S.D. Odintsov and V.K.
Oikonomou).

\end{document}